\documentstyle[12pt]{article}
\def\Re{{\cal R \mskip-4mu \lower.1ex \hbox{\it e}\,}}
\def\Im{{\cal I \mskip-5mu \lower.1ex \hbox{\it m}\,}}
\def\ie{{\it i.e.}}
\def\eg{{\it e.g.}}

\def\etal{{\it et al.}}

\def\sub#1{_{\lower.25ex\hbox{$\scriptstyle#1$}}}
\def\sul#1{_{\kern-.1em#1}}
\def\sll#1{_{\kern-.2em#1}}
\def\sbl#1{_{\kern-.1em\lower.25ex\hbox{$\scriptstyle#1$}}}
\def\ssb#1{_{\lower.25ex\hbox{$\scriptscriptstyle#1$}}}
\def\sbb#1{_{\lower.4ex\hbox{$\scriptstyle#1$}}}

\def\to{\rightarrow}
\def\mh{\ifmmode m\sbl H \else $m\sbl H$\fi}
\def\mch{\ifmmode m_{H^\pm} \else $m_{H^\pm}$\fi}
\def\mt{\ifmmode m_t\else $m_t$\fi}
\def\mc{\ifmmode m_c\else $m_c$\fi}
\def\mz{\ifmmode M_Z\else $M_Z$\fi}
\def\mw{\ifmmode M_W\else $M_W$\fi}
\def\mws{\ifmmode M_W^2 \else $M_W^2$\fi}
\def\mhs{\ifmmode m_H^2 \else $m_H^2$\fi}
\def\mzs{\ifmmode M_Z^2 \else $M_Z^2$\fi}
\def\mts{\ifmmode m_t^2 \else $m_t^2$\fi}
\def\mcs{\ifmmode m_c^2 \else $m_c^2$\fi}
\def\mchs{\ifmmode m_{H^\pm}^2 \else $m_{H^\pm}^2$\fi}
\def\ztwo{\ifmmode Z_2\else $Z_2$\fi}
\def\zone{\ifmmode Z_1\else $Z_1$\fi}
\def\mtwo{\ifmmode M_2\else $M_2$\fi}
\def\mone{\ifmmode M_1\else $M_1$\fi}
\def\tb{\ifmmode \tan\beta \else $\tan\beta$\fi}
\def\xw{\ifmmode x\sub w\else $x\sub w$\fi}
\def\ch{\ifmmode H^\pm \else $H^\pm$\fi}
\def\lum{\ifmmode {\cal L}\else ${\cal L}$\fi}
\def\inpb{\ifmmode {\rm pb}^{-1}\else ${\rm pb}^{-1}$\fi}
\def\infb{\ifmmode {\rm fb}^{-1}\else ${\rm fb}^{-1}$\fi}
\def\epem{\ifmmode e^+e^-\else $e^+e^-$\fi}
\def\ppb{\ifmmode \bar pp\else $\bar pp$\fi}

\newskip\zatskip \zatskip=0pt plus0pt minus0pt
\def\matth{\mathsurround=0pt}

\def\atversim#1#2{\lower0.7ex\vbox{\baselineskip\zatskip\lineskip\zatskip
  \lineskiplimit 0pt\ialign{$\matth#1\hfil##\hfil$\crcr#2\crcr\sim\crcr}}}

\renewcommand{\thefootnote}{\fnsymbol{footnote}}

\begin{document} \begin{titlepage}
\setcounter{page}{1}
\thispagestyle{empty}
\rightline{\vbox{\halign{&#\hfil\cr
&ANL-HEP-PR-93-18\cr
&March 1993\cr}}}
\vspace{1in}
\begin{center}

{\Large\bf
Extraction of Coupling Information From $Z' \to jj$}
\footnote{Research supported by the
U.S. Department of
Energy, Division of High Energy Physics, Contract W-31-109-ENG-38.}
\medskip

\normalsize THOMAS G. RIZZO
\\ \smallskip
High Energy Physics Division\\Argonne National
Laboratory\\Argonne, IL 60439\\

\end{center}

\begin{abstract}

An analysis by the ATLAS Collaboration has recently shown, contrary
to popular belief, that a combination of strategic cuts, excellent mass
resolution, and detailed knowledge of the QCD backgrounds from direct
measurements can be used to extract a signal in the
$Z' \to jj$ channel in excess of $6\sigma$ for certain classes of
extended electroweak models. We explore the possibility that the data
extracted from $Z$ dijet peak will have sufficient statistical power as to
supply information on the couplings of the $Z'$ provided it is used in
conjunction with complimentary results from the $Z' \to \ell^+ \ell^-$
`discovery' channel. We show, for a 1 TeV $Z'$ produced at the SSC, that this
technique can provide a powerful new tool with which to identify the origin of
$Z'$'s.

\end{abstract}


\renewcommand{\thefootnote}{\arabic{footnote}} \end{titlepage}


The observation of a new neutral gauge boson, $Z'$, at the SSC and/or LHC
would be an extremely clear signature for the existence of new physics beyond
the Standard Model(SM). Many analyses have shown{\cite {bigref}} that such
particles are copiously produced at these hadron supercolliders and would be
easily detected via their leptonic decay up to masses of order several TeV for
most extended electroweak model(EEM) scenarios. Of course once the $Z'$ is
detected it would be mandatory to determine its various couplings in order to
determine, if possible, from which EEM it arose. Quite generally, this has
proven to be a more difficult task than one might at first suspect and has
lead to a number of possible approaches advocated in the literature during
the last three years{\cite {idref}}. These various techniques have included
the use of rare decay modes, initially polarized proton beams to produce
new asymmetries, associated production of the $Z'$ together with other gauge
particles, and determining the polarization of final state $\tau's$. All of
these analyses have neglected the possibility that hadronic $Z'$ decays could
be used to obtain coupling information since it is commonly believed{\cite
{lhc}} that the conventional backgrounds($B$) from QCD are so large as to
render this mode unobservable.

Recently, Henriques and Poggioli(HP) of the ATLAS Collaboration{\cite {ATLAS}}
have shown that a combination of strategic cuts, good dijet mass resolution,
and detailed determination of the QCD background in the dijet invariant mass
neighborhood near the $Z'$ can be used to extract a signal($S$) in excess of
$6 \sigma$ at the LHC for a $Z'$ with a mass of 2 TeV with SM-like couplings.
If a determination of the $Z' \to jj$ production cross section,
$\sigma_{jj}$, can
be combined with the corresponding lepton-pair cross section, $\sigma_{\ell}$,
(which is the $Z'$ discovery channel), new information on the $Z'$'s couplings
would be obtained. The purpose of this paper is to analyze this approach for a
1 TeV $Z'$ at the SSC, within the context of several EEM.  (We choose
this mass value in
order to make a comparison with the analyses in{\cite {idref}}.) For this
rather low $Z'$ mass, we find there is sufficient statistics available to
extract coupling information for at least some of these EEM's assuming an
integrated luminosity of $\lum = 10 fb^{-1}$, corresponding to one `standard
SSC' year. Other models, however, lead to too small a value for
${S/{\sqrt {B}}}$ and would require significantly larger values of $\lum$
before useful information could be extracted.

As a first step in this analysis, it is important to remember that the
dominant phase space regions occupied by the QCD and $Z'$ induced dijets
are quite different. Because
of the presence of u- and t-channel poles in the $2 \to 2$ and $2 \to 3$
parton level QCD processes, most of these dijet are at small $p_t$ and prefer
large values of absolute pseudorapidity, $\eta$. Due to the approximate
$1+cos^2 \theta^*$ distribution of the $Z'$ dijets, strong cuts on both the
jet $p_t$ and $\eta$ will substantially increase $S/{\sqrt {B}}$. Of course,
if our cuts are $too$ strong we loose on statistics. As a compromise, for a
1 TeV $Z'$, we employ the cuts $p_t \geq 200$ GeV and $-1 \leq \eta_{j_1,j_2}
\leq 1$ which are similar to those used by ATLAS. These cuts provide us with
jets which are highly isolated and highly central. In order to generate the
QCD background numerically and incorporate, at least approximately, the order
$\alpha_s^3$ corrections{\cite {DKS}}, we employ Next-to-Leading Order(NLO)
 parton
distributions{\cite {pdist}} with a renormalization scale set to minimize
these higher order effects when only the Born-level calculation is performed:
$\mu = M_{jj}/4~cosh(0.7\eta_*)$, with $\eta_*$ being the rapidity of either
jet in the parton center of mass frame. This choice of $\mu$ was explicitly
found to minimize the deviation from the Born calculation due to $\alpha_s^3$
corrections at Tevatron energies by the authors of Ref.{\cite {DKS}}, and also
seems to work just as well as SSC/LHC energies{\cite {Soper}}. Lastly, we
rescale these cross sections by a phenomenological `K-factor' of 1.1. The
result of this
procedure reproduces the order $\alpha_s^3$ results with our kinematic cuts
at the level of 10$\%$, about the same as the uncertainty in the parton
luminosities (see below). (Of course, when the actual procedure we envision is
performed, this QCD background will be $measured$ so that we would no longer
need rely on the perturbative calculations presented here but we can make use
of the data directly. Since no data exists to show how our analysis works,
these best we can do is to use `simulated' data to mimic what we expect to
see at the SSC.)  For a 1 TeV $Z'$, we generate events in the dijet invariant
mass range 0.5 $\leq M_{jj} \leq $ 1.5 TeV subject to the cuts above. Since
dijets involving top(t) quarks may appear somewhat different that those
initiated by the lighter quarks and/or gluons in the final state (since the
top probably decays before fragmentation is complete), we ignore subprocesses
that lead to final state top quarks in this analysis. We note here that the
effect of our cuts is to increase $S/B$ to the level of order 0.1-1$\%$

The dijet signal that arises from the $Z'$ production and decay within a
given model is calculated in the usual manner but also includes a two-loop,
QCD-corrected `K-factor' in
the production process{\cite {Hamberg}} as well as QCD corrections to the
$Z'$ decays to $q {\bar {q}}$ {\cite {QCD}}. These signal events are subjected
to the same cuts as are the QCD background and, of course, only $Z'$ decays to
pairs of lighter quarks are considered. For numerical purposes, we assume a
t-quark mass of 150 GeV, an effective $sin^2 \theta_w$  of 0.2325
{\cite {review}} in EEM couplings, and take $\alpha_s(M_Z)$ =0.117
{\cite {review}}
which we allow to run via the 3-loop renormalization group equations. To
account for the imperfect nature of all supercollider detectors, both the pure
$Z'$ signal as well as the QCD background must be smeared by the dijet mass
resolution. Since the width-to-mass ratio for most $Z'$'s is generally in the
range $0.01 \leq \Gamma_{Z'}/M_{Z'} \leq 0.05$, we anticipate that the dominant
effect of the finite mass resolution will be to smear out the $Z'$ peak and
reduce its statistical significance. For most SSC/LHC detectors, the
anticipated jet energy resolution is expected to be of order{\cite {detec}}
$50 \%/{\sqrt {E}}\oplus 2\%$ which leads to an effective dijet mass
resolution of approximately
$\Delta M_{jj}/M_{jj}=0.034$, as in the ATLAS analysis, which can be seen to
be comparable to the $Z'$ width-to-mass ratio.  We will use this value in our
analysis below. HP have shown in detail how modifying the mass resolution
influences the ratio of $S/B$ and we anticipate that our results would not be
drastically altered if small deviations from our assumed value were to occur.

Once both the QCD and $Z'$ differential cross sections are calculated and
combined using the above description, we integrate it over the allowed ranges
of $\eta$ as well as $cos~\theta^*$ subject to the requirement that the jets
have $p_t \geq 200$ GeV. This result is further integrated over dijet
invariant mass bins of width 25 GeV, which is comparable to the resolution for
dijets with pair masses near $M_{Z'}$, and multiply by the integrated
luminosity; Gaussian statistical fluctuations are included for each mass bin.
At this point, all we have succeeded in doing is generating a `set of data' to
which rather strict cuts have been applied. As one might expect, a plot of
this `data' would show no apparent structure in the neighborhood of 1 TeV as
we will see below.

We now deviate a bit from the HP analysis and make use of the fact that the
mass and width of the $Z'$ will already be determined with reasonably high
precision from the dilepton data {\it {before}} we go hunting in dijets,
 \ie,~we assume that the $Z'$ has already been discovered. The reason that
this is important for our dijet analysis is that the leptonic data tell us
{\it {where}} to look in dijet invariant mass and the approximate size of the
`signal region'. Since we want this region to be at least $\pm 2\Gamma_{Z'}$
wide(roughly speaking) and we are assuming a $M_{Z'}$ of 1 TeV, we will define
the signal region to be within 100 GeV of 1 TeV for all EEM's. We now look at
our `data' outside of this signal region; we find it convenient at this point
to introduce the dimensionless variable $x_{jj}=M_{jj}/M_{Z'}$. One finds that
by rescaling our `data' by $x_{jj}^5$, the resulting distribution becomes
reasonably flat in $x_{jj}$ except for the range $0.5 \leq x_{jj} < 0.7$ where
the effects of the strong $p_t$ and $\eta$ cuts become noticeable. Because
we need
to determine the background as precisely as possible outside the signal region,
we do not include this $x_{jj}$ range in out fit. Next, we take our rescaled
`data' and fit to a polynomial outside the signal region; our best
$\chi^2/d.o.f.$ (which is EEM
dependent but quite close to unity in all cases) results for a fit to a
polynomial of
degree 7 whose coefficients (and associated errors) are obtained by
least-squares using the singular value decomposition technique in the usual
way. The use of a polynomial of a larger degree does not result in an
improvement in our $\chi^2/d.o.f$. Next,
we extrapolate into the signal region using this polynomial fit and subtract
our QCD background which, hopefully, will result in a dijet $Z'$ `peak'
provided the $Z'$'s couplings are sufficiently strong and enough statistics
are available.

At this point we've rendered the $Z'$ bump in dijets `visible' and we fit the
peak to a Gaussian and/or a relativistic Breit-Wigner. In either case, we let
the amplitude, width, and position of the maximum float and obtain a best fit
from a second $\chi^2$ analysis. We then can either integrate under
the fitted peak, use the narrow-width approximation, or simply count the
excess of events in the signal region to obtain the total number of
$Z'\to jj$ events.

Even if the $Z'$ can be observed in the dijet channel, we cannot use our
determination of the integrated cross section directly to obtain coupling
information for a number of reasons. ($i$) As an absolute number of events,
our result suffers from a number of systematic uncertainties, \eg,~variations
in the structure functions and machine integrated luminosity. ($ii$) Our
observable depends explicitly on the width of the $Z'$, \ie,~it depends on
what final states are allowed in $Z'$ decay. To alleviate such problems we
propose to take the ratio of the number of fitted $Z'$-induced dijet events to
the number of $Z'$-induced dileptons. This quantity, $R$, essentially measures
the ratio $\sigma_{jj}/\sigma_{\ell}$, subject to the various cuts, is
independent of luminosity uncertainties, and as we will see is extremely
weak in its dependence on the choice of structure functions. To be specific,
we define the lepton-pair cross section, $\sigma_{\ell}$, to include a cut of
$-2.5 \leq \eta_{\ell} \leq 2.5$
on both outgoing leptons. Except for the various cuts then, $R$ tells us the
ratio of the dijet to leptonic widths of the $Z'$ in the narrow width
approximation.

We should, of course, convince the reader that $R$ is {\it {worth}}
determining, \ie,~that it is sensitive to the $Z'$ couplings. To this end we
must examine a number of specific EEM's; we consider four representative
examples in what follows. ($i$) The Left Right Symmetric Model(LRM)
{\cite {LRM}} with the ratio of right-handed to left-handed couplings,
$\kappa =g_R/g_L$, set to unity; ($ii$) The Alternative version of the LRM,
which we denote by ALRM
{\cite {ALRM}}; ($iii$) The $E_6$ rank-5 models(ER5M){\cite {pr}}, which
contains a free parameter, $-\pi/2 \leq \theta \leq \pi/2$, which determines
all of the $Z'$ couplings to fermions; and ($iv$) a $Z'$ with the same
couplings as the $Z$ of the SM, which we call SSM. This list is far from
exhausting the set of models on the market at present. Fig.~1a shows the ratio
$R$ as a function of $\theta$ in the ER5M case for two different structure
function choices. We see that $R$ is quite sensitive to $\theta$ but that
possible deviations due to structure functions cancel almost entirely
in taking the ratio;
this has been confirmed numerically by examining the results obtained through
the use of several other structure function sets. Fig.~1b shows the
corresponding
dependence of the ratio $R$ on the parameter $\kappa$ in the LRM; we again see
that $R$ is quite sensitive to variations in the fermionic couplings and very
insensitive to the choice of parton densities. (We again remind the
reader that this ratio involves cross sections to which the above cuts have
been applied and where top-quark final states are ignored.) For the LRM with
$\kappa =1$(ALRM, SSM) we obtain $R= 30.5(3.92, 18.9)$ for
the MTS1 set of parton densities of Morfin and Tung{\cite {pdist}} which we
now take as our default. Clearly, the values of $R$ range over
more than an order of
magnitude for the various models we've considered thus demonstrating its
sensitivity to the possible $Z'$ couplings. We have found that the
range of allowed values for
$R$ in other models could be much greater. If $R$ can be reliably determined
by using the dijet data we will have a new piece of the $Z'$ coupling puzzle.
It remains to be seen whether we can in fact perform this feat.

Let us use the LRM as a test case. Fig.~2a shows the number of dijet events,
$N_{jj}$, in each
25 GeV mass bin as a function of $x_{jj}$ for our range of interest; the
errors are contained within the crosses and no apparent evidence for a $Z'$ is
yet visible. We note, as discussed above, that
for $x_{jj}\geq 0.7$ the distribution has a positive second derivative but
that this begins to change below $x_{jj}\simeq 0.7$ due to our cuts. Rescaling
$N_{jj}$ by $x_{jj}^5$ (and an overall trivial constant) leads to Fig.~2b
where we see that this rescaled distribution, $N^0_{jj}$, differs
significantly from unity only for small $x_{jj}$ and
is only weakly $x_{jj}$ dependent otherwise as advertised. (We note that
modifying the power of
$x_{jj}$ in the rescaling procedure by a small amount in an attempt to
further flatten this distribution will not effect the results of our fit
since the `gross' $x_{jj}$-dependence has already been accounted for.) We see
that this rescaling has finally allowed the statistical fluctuations in the
data to become visible for the first time although the $Z'$ still remains
hidden.
Now the values of $N^0_{jj}$, together with the associated errors, are fit
outside the signal region by
our degree 7 polynomial, extrapolated into the range $0.9 \leq x_{jj} \leq
1.1$ and subtracted. The result is shown in Fig.~3a which displays the number
of excess dijet events, $N^{exc}_{jj}$, as a function of $x_{jj}$ as well as
the best fit of this excess to a Gaussian(G) and a relativistic Breit-Wigner
(BW), as described above. Our result is quite similar to that obtained in the
ATLAS analysis even though the effects of particle fragmentation and detailed
detector properties have not been included in the present analysis. For the
BW(G) case, the fitted peak is located at
995(994) GeV with a width of 65(34) GeV; the actual width of this $Z'$ used
as input into our analysis is 20.6 GeV. Thus we see that the effect of the
finite mass resolution and fitting procedure is to broaden the peak (as well
as flatten it). If we follow this identical procedure for the SSM case, we
arrive at Fig.~3b. Here, for the BW(G) fit we obtain a peak position of 999
(1001) GeV with a width of 57(31) GeV, the actual input value being 30.2 GeV.
Since the underlying parton-level process is of the BW type, we might be
somewhat biased in
favor of this particular choice. We find, however, that the Gaussian fit has a
slightly better $\chi^2$.

Knowing the number of $Z'$-induced lepton pair events from our previous
analysis{\cite {JLHTGR}}, we can sum the appropriate excess event
distributions and
arrive at the extracted values for the ratio $R$ for both the LRM and SSM
cases. By just counting the excess events, we obtain $R_{LRM} =40.7\pm 4.6$
and $R_{SSM} =22.5\pm 2.4$,
both of which are quite close to the theoretical expectations above. (Note the
tendency of these values to lie on the high side of the actual expected
values.) In the LRM case, if we force a fit to the model parameter space,
this result allows us to place a constraint on the value of
the parameter $\kappa$ by comparison theoretical expectations for this model
shown in Fig.~1b. At the $3\sigma$ level, we learn that $0.83 \leq \kappa \leq
1.11$. If, instead, we integrate the number of events under the {\it {fitted}}
peak, we find a somewhat smaller pair of results: $R_{LRM}=34.9 \pm 4.0$ and
$R_{SSM}=20.4 \pm 2.2$ for the Gaussian fit. For the LRM, this results in a
corresponding bound with essentially the same range as above but now at the
$95\%$ CL.

Before continuing, we would again like to stress the important role played
by the {\it {leptonic}} data in this analysis. ($i$) The leptonic data has
told us where to look in dijet invariant mass for the $Z'$ and has given us
its approximate width (after de-convoluting the dilepton pair mass
resolution). ($ii$) The leptonic data allows us to normalize our dijet
results which removes luminosity and structure function uncertainties and
provides us with a new observable, which is independent of other
potentially exotic modes
that the $Z'$ might possess, and is highly sensitive to the $Z'$'s couplings.

What happens if we perform the analysis above for the other models?
Fig.~4 shows the
predicted number of $Z'$-induced dijet events in the mass bin containing the
$Z'$ peak as a function of the parameter $\theta$ in the ER5M. As can be easily
seen, for all of these models this value is at least 4-5 times smaller than
those for the LRM and SSM shown in the fits in Fig.~3. In fact, for all values
of $\theta$, this `excess' is below $\simeq 1.6\sigma$ in significance and is
very probably unobservable at this level of integrated luminosity. A similar
situation arises in the ALRM case where the bin with the $Z'$ dijet peak
contains only 3044 events, approximately a factor of 7 below the LRM case,
and is comparable to that found for the ER5M.

To show what happens for these less fortunate cases, less us consider the ER5M
with $\theta=-\pi/2$, which is referred to as `model $\chi$' in the literature.
We choose this case as it essentially has the largest number of excess
events expected in
the bin containing the $Z'$ peak of all the ER5M's and also predicts
a value larger
than the ALRM case. In other word, if we cannot extract a signal for model
$\chi$, we cannot do it for {\it {any}} ER5M or for the ALRM. Following the
same procedure as above leads    to the excess event distribution shown in
Fig.~5a. We see {\it {no}} apparent excess of any significance in the region
near $x_{jj}=1$. (Of course, we can always try and {\it {force}} a BW or G fit
in this region but the result would have a terribly bad $\chi^2$.) Thus we
conclude that $Z'$'s from the ER5M or ALRM would not be visible in dijets
without significantly more integrated luminosity. To achieve the same
$S/{\sqrt {B}}$ as in either the SSM or LRM cases would require an increase in
$\lum$ by at least a factor of 25; of course such a large value is not
necessary to render the $Z'$ dijet peak visible. To demonstrate this, we show
in Fig.~5b the number of excess events in dijets for model $\chi$ again, but
with a factor of ten increase in integrated luminosity, $100 fb^{-1}$.
(We remind the reader
that we are aided by the fact that we know from the dilepton data the
approximate location of the peak.) Now that $Z'$ peak is `visible' after
background subtraction, we can perform a fit, shown in Fig.~5b, as in the SSM
and LRM cases and extract a value by simply counting the number of excess
events, $R_{\chi}=12.7 \pm 2.7$, to be compared with the
theoretical value of 9.6. (We again see that the value we extract tends to be
systematically high.) Furthermore, at the $95 \%$ CL, we find that even
this meager data disallows values of $\theta$ between $12^{\circ}$ and
$36^{\circ}$.~If, instead, we integrated under the Gaussian fit we would obtain
a similar result: $R_{\chi}=12.2 \pm 2.6$, with a similar range of $\theta$
now being disallowed at $95 \%$ CL: $9^{\circ} \leq \theta \leq 39^{\circ}$.

A summary of the analysis presented here is as follows:

($i$) We have shown that the $Z' \to jj$ peak can be observable with high
significance, for a 1 TeV $Z'$, at the SSC given an integrated luminosity of
$10 fb^{-1}$ at least for some classes of extended electroweak models. It
would appear that more massive $Z'$'s or $Z'$'s of the same mass but arising
from other models might also be observable provided sufficient integrated
luminosity were available. We consider the fact that the $Z'$ will already be
known to exist from the dilepton data, with a given mass and width, to be of
extreme importance in this type of analysis. The value of
tight $p_t$, $\eta$, and dijet mass cuts, together with excellent dijet mass
resolution, was also shown to be of great significance is obtaining
a dijet data set with which to explore $Z'$ properties. The importance of
obtained detailed data on the QCD background away from the the resonance
region so that a precise background subtraction can be performed cannot be
overly emphasized.

($ii$) If sufficient statistics become available, the number of excess dijet
events remaining after QCD background subtraction can be used, in combination
with the leptonic data, to determine the ratio $R$. $R$ was seen to be quite
sensitive to model couplings yet insensitive to parton structure function,
luminosity, and $Z'$ decay mode uncertainties. $R$ was shown to be capable or
restricting the allowed range of the parameter spaces of both the LRM and the
ER5M. When combined with other probes of the $Z'$ fermionic couplings, a
complete determination might now be obtainable.

($iii$) One may possibly be able to extend the procedure developed here to
other $Z'$ decay modes provided tagging techniques with high efficiencies are
found.

Hopefully, the $Z'$ will be there for us to explore.

\vskip.25in
\centerline{ACKNOWLEDGEMENTS}

The author would like to thank D. Soper for his assistance in the estimates
of the order $\alpha^3_s$ dijet production cross-section at the SSC used
in the above and
L. Poggioli, of the ATLAS Collaboration, for providing the details of
their $Z' \to jj$ analysis. The author would also like to thank J.L. Hewett
and P.K. Mohapatra for discussions related to the present work.
This research was supported in part by the U.S.~Department of Energy
under contract W-31-109-ENG-38.

\newpage

%
\def\MPL #1 #2 #3 {Mod.~Phys.~Lett.~{\bf#1},\ #2 (#3)}
\def\NPB #1 #2 #3 {Nucl.~Phys.~{\bf#1},\ #2 (#3)}
\def\PLB #1 #2 #3 {Phys.~Lett.~{\bf#1},\ #2 (#3)}
\def\PR #1 #2 #3 {Phys.~Rep.~{\bf#1},\ #2 (#3)}
\def\PRD #1 #2 #3 {Phys.~Rev.~{\bf#1},\ #2 (#3)}
\def\PRL #1 #2 #3 {Phys.~Rev.~Lett.~{\bf#1},\ #2 (#3)}
\def\RMP #1 #2 #3 {Rev.~Mod.~Phys.~{\bf#1},\ #2 (#3)}
\def\ZP #1 #2 #3 {Z.~Phys.~{\bf#1},\ #2 (#3)}
\def\IJMP #1 #2 #3 {Int.~J.~Mod.~Phys.~{\bf#1},\ #2 (#3)}

\newpage

%
{\bf Figure Captions}
\begin{itemize}

\item[Figure 1.]{Predicted values of $R$ in the (a)ER5M as a function of
$\theta$ and in the (b)LRM as a function of $\kappa$. The solid curve is
the result of using the Morfin-Tung MTS1 parton densities while the dashed
curve is the result of using the HMRSB densities of Harriman \etal .}
\item[Figure 2.]{(a) The number of dijet events in 25 GeV wide
mass bins produced at
the SSC as a function of $x_{jj}$ in the invariant mass region near the 1 TeV
$Z'$ of the LRM after the cuts described in the text have been applied. An
integrated luminosity of 10 $fb^{-1}$ has been assumed. The errors are totally
contained within the crosses. (b) Same as (a), but rescaled by a factor of
$x_{jj}^5$ and an overall trivial constant.}
\item[Figure 3.]{Invariant mass distribution, in 25 GeV wide bins, of
the excess
dijet events due to the $Z'$ of the (a)LRM and (b)SSM after QCD background
subtraction at the SSC assuming the same integrated luminosity as in Fig.~2.
The solid(dash-dotted) curve is the result of performing a best
fit to the excess assuming a Gaussian(Breit-Wigner) shape for these events.}
\item[Figure 4.]{Predicted number of {\it {signal}} dijet events at the SSC
in the 25 GeV wide invariant mass bin containing the $Z'$ peak, assuming an
integrated luminosity of $10 fb^{-1}$, in the ER5M as a function of $\theta$.}
\item[Figure 5.]{Same as Fig.~3, but for the ER5M $\chi$ assuming an
integrated luminosity of (a) $10 fb^{-1}$ and (b) $100 fb^{-1}$. In the second
case, both Gaussian(solid) and Breit-Wigner(dash-dotted) fits to the peak are
also shown.}
\end{itemize}

\end{document}